\documentclass[a4paper,fleqn,usenatbib]{mnras}
\usepackage{ae,aecompl}
\usepackage{amsmath,amssymb,graphicx}
\usepackage{bm}
\usepackage{color}
\usepackage{xcolor}
\usepackage{times}
\usepackage{microtype}
\usepackage{booktabs}
\usepackage[normalem]{ulem}
\usepackage[varg]{txfonts}
\usepackage{multirow}
\usepackage{relsize}
\usepackage{widetext}

\newcommand{\beq}{\begin{equation}}
\newcommand{\eeq}{\end{equation}}
\newcommand{\MM}{\mathcal{M}}

\title[Redshift evolution of double neutron star mergers]{Imprints of the redshift evolution of double neutron star merger rate on
	the signal to noise ratio distribution}

\author[Shilpa Kastha et al.]{
	Shilpa Kastha$^{1,2}$\thanks{E-mail: shilpakastha@imsc.res.in},
	M. Saleem$^{3}$\thanks{E-mail: msaleem@cmi.ac.in},
	K. G. Arun$^{3,4}$\thanks{E-mail: kgarun@cmi.ac.in}
	\\
	$^{1}$The Institute of Mathematical Sciences, IV Cross Street, CIT Campus, Taramani, Chennai 600113. \\
	$^{2}$Homi Bhabha National Institute, Training School Complex, Anushakti Nagar, Mumbai, 400094, India.\\
	$^{3}$Chennai Mathematical Institute, Siruseri, 603103 Tamilnadu. \\
	$^{4}$Institute for Gravitation and the Cosmos, Pennsylvania State University, State College, PA 16802.\\
	}

\date{Accepted XXX. Received YYY; in original form ZZZ}

\pubyear{2019}

\begin{document}
\label{firstpage}
\pagerange{\pageref{firstpage}--\pageref{lastpage}}
\maketitle

\begin{abstract}
	Proposed third generation gravitational wave (GW)
interferometers such as Cosmic Explorer will have the sensitivity to
observe double neutron star (DNS) mergers up to a redshift of $\sim 5$
with good signal to noise ratios. We argue that the {comoving spatial
distribution of DNS mergers leaves a unique imprint on the statistical
distribution of signal to noise ratios (SNRs) of the detected DNS
mergers}. Hence the SNR distribution of DNS mergers will facilitate a
novel probe of their redshift evolution independent of the luminosity
distance measurements. We consider detections of DNS mergers by the
third generation detector Cosmic Explorer and study the SNR distribution
for different possible redshift evolution models of DNSs and employ
Anderson Darling  \textit{p-value} statistic to demonstrate the
distinguishability between these different { models}. We find that a few
hundreds of DNS mergers in the Cosmic Explorer era
 will allow us to distinguish between different models of redshift
evolution. We further apply the method for various SNR
distributions arising due to different merger delay-time and star
formation rate (SFR) models and show that for a given SFR model, the SNR distributions are sensitive to the delay time distributions. Finally, we investigate the effects of sub-threshold events in distinguishing between different merger rate distribution models.
\end{abstract}


\section{Introduction}
	The first two observation runs of advanced LIGO and Virgo
interferometers have led to the detections of 10 binary black hole
mergers~\citep{Discovery,GW151226,GW170104,GW170814,GW170608,GWTC-1,GW-Catalogue2,
BBH1strun}  and a binary neutron star merger~\citep{GW170817}. The
binary neutron star merger was also observed in various bands of the
electromagnetic spectrum from gamma rays to the
radio~\citep{MMA,Goldstein:2017mmi,KN-discovery-170817,Ruan:2017bha,Margutti:2018xqd,DAvanzo:2018zyz,Troja:2018ruz,Lyman:2018qjg,Resmi:2018wuc,Lazzati:2017zsj},
These detections have given us unique insights about the
astrophysics~\citep{Astro,Rates,PE,GW-GRB,Multi-messenger2017,Kilonova_GW170817,
Progenitor-BNS,Neutrinos-BNS,BNS-properties19},
cosmology~\citep{LVCHubble} and fundamental
physics~\citep{TOG,GW151226,GW170104,GW-GRB,BNS-radii19,BNS-properties19,Multi-messenger2017,Kilonova_GW170817,
Progenitor-BNS,Neutrinos-BNS,BNS-Postmerger}. With the planned upgrades
of advanced LIGO and other similar interferometers [Virgo~\citep{VIRGO},
KAGRA~\citep{KAGRA_ref}, LIGO-India~\cite{Ligo-india}] joining the
world-wide network of GW detectors, we are gearing up for exciting times
in GW astronomy. There are ongoing research and development activities
towards third generation ground-based detectors such as Einstein
Telescope (ET)~\citep{ET:Sathya,ET} and Cosmic Explorer~\citep{CEDwyer}.
Following the success of LISA Pathfinder~\citep{LPF}, the space-based LISA mission is now funded~\citep{Babak2017}. With these developments, GW astronomy is going to be a very active field of research in the coming decades~\citep{SathyaSchutzLivRev09}. 

  One of the key science goals of third generation detectors is to
probe the redshift evolution of compact binary mergers in the
universe~\citep{ArunET09,Broeck13}.
The redshift evolution of compact binaries would shed light on several
unknown facets of star formation and stellar evolution~\citep{Vitale:2018yhm}. The conventional
way to reconstruct the redshift evolution of compact binaries is by
measuring the luminosity distance to each merger and use a cosmological
model to translate the distance estimate to redshift. By repeating this
for all the detected events, one can reconstruct the redshift evolution
for compact binary mergers~\citep{Broeck13,FishbachHolz2018}.
However, luminosity distance is
one of the 14 parameters that describes the compact binary merger and
hence the parameter estimation of each event involves multi-dimensional
maximization of the likelihood function. This  is computationally very
demanding for third generation detectors because of  their low frequency
sensitivity due to which the signals will last for several minutes to
hours in the sensitivity band of the detector. This problem is even
worse for low mass systems, such as binary neutron star mergers, which
merge in the kilo Hertz frequencies. As third generation detectors are expected to detect tens of
thousands of compact binary mergers, parameter estimation is going to
pose challenges in terms of computing resources. { {Hence,
detailed parameter estimation follow-up of GW events will be prioritized
based on their astrophysical significance and low-significant events
(also known as sub-threshold events) are most likely not to be followed
up. More over, \cite{Huang:2018tqd} showed that the parameter estimation of low-significant events are subjected to noise systematics and often do not yield reliable estimates.}}

{ {On the other hand, low-significant events, if astrophysical in origin, are mostly likely to be either from higher redshifts or a near edge-on system. A population with high redshift events is best suited for probing the redshift evolution. In some cases, such low significant events might be identified to be in spatial and/or temporal coincidence with triggers from electromagnetic search which may boost the significance of being astrophysical in origin. Therefore, one may not want to ignore the low-significant events completely.
Hence developing
methods to probe the redshift evolution of compact binary mergers which
does not rely on the parameter estimation is going to be very useful. 
Here we propose a novel method which can be used
to probe redshift evolution of binary neutron star mergers by using only
the signal to noise ratio of each event. This
information is likely to be readily available as outcomes of detection
algorithms, irrespective of whether the detailed parameter estimation follow-up of the signal is performed or not.}}


Recently Schutz~\citep{Schutz2011} pointed out, on very general
grounds, that the observed signal to noise ratio (SNR) of the GW events
detected by GW detectors should follow a universal distribution
if the underlying spatial density of the source population is
constant within the 
volume accessible to the detectors. This
distribution is independent of the type of the sources and hence
referred to as universal. As argued by Schutz~\citep{Schutz2011}, the
universality arises from the fact that the SNR of the GW events are
inversely proportional to the luminosity distance ($\rho\propto
\frac{1}{D_L}$) and at relatively low redshifts (say, $z\lesssim 0.1$) the
luminosity distance and co-moving distance can be approximated to be the
same. More precisely, following Chen and Holz~\citep{ChenHolz2014}, the
probability of a source (say a compact binary merger) to be found within
a shell of thickness $dD$, at a co-moving distance of $D$, goes as $f_D
dD\propto D^2\,dD$, if the co-moving number density of the source
population is constant. Since $\rho\propto \frac{1}{D}$, the distribution
of SNR corresponding to the particular source distribution, can easily
be shown to follow $p(\rho)=f_D\big|\frac{dD}{d\rho}\big|\propto \frac{1}{\rho^4}$. After normalization, we obtain
\beq
p(\rho)=\frac{3\rho_{th}^3}{\rho^4},\label{snrdist}
\eeq
where $\rho_{th}$ is the SNR threshold used for detection. The
above derivation crucially assumes that the properties of the source
population (such as mass distribution) do not evolve with redshift. Chen
and Holz~\citep{ChenHolz2014} explored various implications of this
universal distribution for the sources detectable by second generation
detectors such as advanced LIGO/Virgo. This universal distribution is
also an ingredient  used in~\citep{Rates} to derive a bound on the rate of the binary
black hole mergers from the first observation run of LIGO~\citep{GW151226}.

	Motivated by \citep{Schutz2011} and \citep{ChenHolz2014}, in this paper, we study the SNR distribution  of compact binary mergers but for cosmological sources. For binary black hole mergers, their mass distribution is likely to influence the SNR distribution as much as the cosmological evolution ({ see for instance  \cite{Vitale2016}}) which makes it difficult to disentangle the two effects. {However, that is not the case with DNS mergers as the masses are expected to vary over a relatively smaller range compared to binary black hole mergers.} The planned Cosmic Explorer (CE) will have enough sensitivity to  detect DNS mergers with good SNR ($\sim 12$) up to a redshift of $\sim 5$~\citep{ET:Sathya, ET}. 

Considering CE sensitivity as representative of the third
generation GW detector, in
this paper, we study how the SNR distribution of DNS mergers observed
by CE gets affected  by the redshift evolution of their  rate
density and hence use the detected SNR distribution to probe the
underlying redshift evolution of DNS mergers. Considering
astrophysically motivated models for the redshift evolution of DNS
merger rate density, we study how distinguishable are the resulting SNR
distributions from each other. 
 We find that observations of the order
of a few hundreds of DNS mergers are sufficient to distinguish between
different redshift evolution models. {As the projected detection rate of
DNS mergers by the third generation GW detectors is of the order a few hundreds to thousands~\citep{ArunET09},
 one year of observation by CE may itself be
sufficient to track the redshift evolution of DNS using this method.

The paper is organized in the following way. In Sec.~\ref{sec.2} we
consider the cosmological effects on the optimal SNR of compact binary
sources. In Sec.~\ref{sec.3}
we explore how the different DNS merger rate densities affect the
SNR distributions. In Sec.~\ref{Statistical_tests} we discuss whether the
distributions corresponding to all the merger rate densities are
distinguishable from each other. 

\section{Effects of cosmological expansion on the signal to noise
ratio of compact binaries}\label{sec.2}

	The data analysis technique of matched filtering~\citep{Helstrom68,Wainstein} is usually employed to detect compact binary mergers using GW observations. Matched filtering involves cross-correlating various copies of the expected gravitational waveforms (templates) corresponding to different signal parameters (such as masses and spins), with the data, which potentially contains the signal (in addition to the noise). The template which maximizes the correlation is referred to as optimal template and the corresponding signal to noise ratio is called optimal SNR which is defined as 
	\beq
		\rho=\sqrt{4\int_0^\infty\frac{|\tilde{h} (f)|^2}{S_h(f)}df},\label{optsnrdef}
	\eeq
	where $S_h(f)$ is the detector's power spectral density (PSD) and $\tilde{h} (f)$ is the frequency domain gravitational waveform (See, for instance, Sec.~(5.1) of \citep{SathyaSchutzLivRev09} for details). 

	We employ restricted post-Newtonian (PN) waveform (RWF),
$\tilde{h}(f)=\mathcal{A}f^{-7/6}e^{i\psi(f)}$, where $\mathcal{A}$ is
the amplitude and $\psi(f)$ is the frequency domain GW phase. In RWF,
the PN corrections to the amplitude of the gravitational waveform are
ignored but the phase is accounted for to the maximum accuracy. Using
the RWF, the optimal SNR for GW events of compact binary systems can be
expressed as~\citep{CF94}
	\begin{widetext}
    \begin{eqnarray}
		\rho(m_1,m_2,D_L,\theta,\phi,\psi,\iota)&=& 
		\sqrt{4\frac{\mathcal{A}^2}{D_L^2}\left[F_+^2(\theta,\phi,\psi)(1+\cos^2\iota)^2+4F_\times^2(\theta,\phi,\psi)\cos^2\iota\right]I(M)},\label{fullsnr}
	\end{eqnarray}
	\end{widetext}
where $F_{+,\times}(\theta,\phi,\psi)$ are the antenna pattern functions
for the `plus' and `cross' polarizations,
$\mathcal{A}=\sqrt{5/96}\;\pi^{-2/3}\MM^{5/6}$, $\MM$ is the chirp mass,
which is related to the total mass $M$ by $\MM=M\,\eta^{3/5}$, where $\eta=\frac{m_1m_2}{M^2}$ is the symmetric mass ratio of the system and $m_1$, $m_2$ are the component masses.  { The four angles $\{\theta,\phi,\psi,\iota\}$ describe the  location and orientation of the source with respect to the detector.} $I(M)$ is the frequency integral defined as
	\beq
		I(M)=\int_0^{\infty}\frac{f^{-7/3}}{S_h(f)}df\simeq\int_{f_{low}}^{f_{\rm LSO}}\frac{f^{-7/3}}{S_h(f)}df\label{freq_integral}.
	\eeq
	In the last step, we have replaced the lower and upper limit of the integral by the seismic cut off, $f_{low}$, of the detector and the frequency at the last stable orbit of the black holes with masses $m_1$ and $m_2$, respectively. The GW frequency at the last stable orbit (LSO) upto which PN approximation is valid, as a function of the total mass M is, $f_{\rm LSO}=\frac{1}{6^{3/2}\,\pi\,M}$. This is the expression for the frequency at the LSO of a Schwarzschild BH with total mass $M$.

	As we use CE as a proxy for third generation detectors in this
work, the signal to noise ratio computations uses the following fit for the Cosmic explorer wide band (CE-wb) sensitivity curve~\citep{CE_WB}
	\begin{eqnarray}
		S_h(f) &=& 5.62\times10^{-51} + 6.69\times10^{-50}f^{-0.125} + 7.80\times10^{-31} f^{-20}  \nonumber \\
		&+& 4.35\times10^{-43} f^{-6} + 1.63\times10^{-53}\,f + 2.44\times10^{-56}\,f^2 \nonumber \\
		&+& 5.45\times10^{-66}\,f^5\,\, {\rm Hz^{-1}}\,.
	\end{eqnarray}
Next we discuss the effect of cosmology on the gravitational waveform
and hence the expression for SNR.

\subsection{Effects of cosmological expansion}

	Assuming a flat $\Lambda$ CDM cosmological model~\citep{Planck2013,Planck2018} for the universe, we 	explore the modification to the SNR for compact binary systems at cosmological distances. Cosmological expansion of the universe affects the gravitational waveforms in two ways. According to general relativity, GW amplitude is inversely proportional to the co-moving distance $D$, which is no longer same as the luminosity distance $D_L$ but is related by $D_L=D\,(1+z)$, where $z$ is the redshift to the source. Due to the expansion of universe, there is an additional factor $(1+z)$ in the denominator of the amplitude. This is already accounted in Eq.~\ref{fullsnr} as it is written in terms of $D_L$. 
Secondly, due to the cosmological expansion, the gravitational wave
frequency gets redshifted. This results in redefining the chirp mass in
such a way that the observed chirp mass $(\MM)$ is related to the
corresponding chirp mass in the source frame $(\MM_{\rm source})$, by
$\MM=\MM_{\rm source}\,(1+z)$. This happens due to the fact that the
only time scale of the problem $\frac{G\MM}{c^3}$, is redshifted, which
is completely degenerate with mass and hence leading to the notion of
redshifted mass (see Sec.~4.1.4 of \citep{Maggiore}). In order to
explicitly incorporate these effects, we re-write the expression for SNR in Eq.~\ref{fullsnr} as
	\beq
	\rho=\frac{\MM_{\rm
	source}^{5/6}}{D(1+z)^{1/6}}\;g(\theta,\phi,\psi,\iota)\;\sqrt{I(M)},\label{simplesnr}
	\eeq
	where all the angular dependencies in the waveform are captured into the definition of $g(\theta,\phi,\psi,\iota)$ and other variables have their usual meanings.

	In a flat FLRW cosmology, the comoving distance (following $c=G=1$ units), corresponding to a redshift z (Ref.~\citep{Hogg1999,Wright:2006}) is given by
	\beq
		D(z)=\frac{1}{H_0}\int_0^z\frac{dz^\prime}{E(z^\prime)}\label{DL},
	\eeq
	where $H_0$ is the Hubble constant and
	\beq
		E(z)=\sqrt{\Omega_m(1+z)^3+\Omega_\Lambda},
	\eeq	
	with the total density parameter ($\Omega$) consisting of matter (dark and baryonic) density  ($\Omega_m$)  and cosmological constant ($\Omega_\Lambda$). Throughout this work, we consider a cosmology with $\Omega_\Lambda=0.7$ and $\Omega_m=0.3$ and $H_0=72 {\rm km/Mpc/sec}$~\citep{Planck2013,Planck2018}.  

	Given that $z$ is a function of $D$ (Eq.~\ref{DL}), it is
clear from Eq.~\ref{simplesnr}  that the simple scaling relation for
SNR ($\rho \propto 1/D$) would no longer hold. Hence it is obvious that
the universal SNR distribution given in Eq.~\ref{snrdist} does not
apply any more. In the next section we discuss the effect of redshift evolution
on the SNR distribution and how the SNR distribution encodes the
signature of merger rate density evolution.
 	
\section{ Imprints of co-moving merger rate density evolution of
DNS systems on the SNR distribution}\label{sec.3}

	\begin{figure*}
		\centering
	    \includegraphics[scale=0.5]{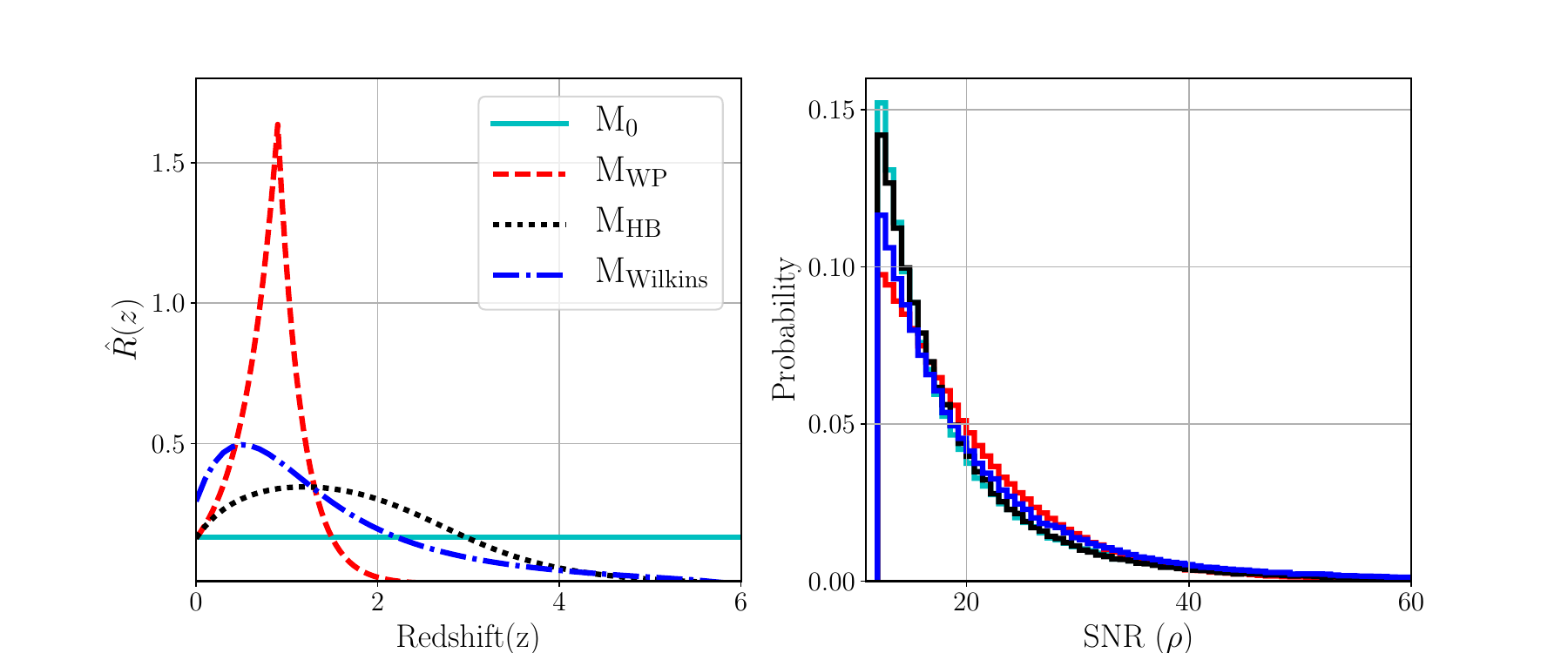}
		\caption{Figure on the left panel shows the evolution of
co-moving merger rate density with redshift for four different models,
$M_0$ stands for the constant comoving merger rate, $M_{\rm{WP}}$
represents the model for rate density evolution obtained by Wandermann
\& Piran~\citep{WP15}, $M_{\rm{HB}}$ and $M_{\rm{Wilkins}}$ denote the merger rate models obtained in Ref.~\citep{Regimbau09, Broeck13} following the star formation rates given in ref.~\citep{Hopkins_Beacom06} and \citep{Wilkins08} respectively. Figure on the right most panel contains the corresponding normalised SNR distributions.}\label{fig:various}
	\end{figure*}


	Usually it is assumed that the DNS formation rate follows the star formation rate while their merger rate will depend also on the delay time distribution: the distribution of the time delay between the formation and the merger.
	Hence following Ref.~\citep{Regimbau09}, binary merger rate density can be written as
	\begin{equation}
		R(z) \propto \int_{t_d^{\rm min}}^\infty \frac{\dot{\rho}_\ast(z_f(z,t_d))}{1+z_f(z,t_d)}P(t_d)\,dt_d,
		\label{eq;Rz}
	\end{equation}
	where $\dot{\rho}_\ast$ is the star formation rate, {$t_d$ the delay time and $t_d^{\rm min}$ the minimum delay time for a binary to merge since its formation}. The redshift z describes the epoch at which the compact binary merges where as $z_f$ is the redshift at which its progenitor binary forms and they are related by a delay time $t_d$. The factor $P(t_d)$ is the distribution of the delay time. According to various population synthesis
	models~\citep{Ando_2004,OShaughnessy07,deFreitasPacheco:2005ub,
	Tutukov94, Lipunov95, Belczynski_2008}, the delay time follows a power-law distribution, $P(t_d)\propto 1/t_d$, {with $t_d>t_d^{\rm min}$}. The factor $(1+z_f)^{-1}$ takes into account the cosmological time dilation between the star formation and the merger. 

	For our analysis in this paper, we use two merger rate models,
following the two star formation rate models proposed by
\cite{Hopkins_Beacom06} and \cite{Wilkins08} and denote them as $M_{\rm
HB}$ and $M_{\rm Wilkins}$, respectively. In both the cases we
consider~\citep{Belczynski_2008, Belczynski00} $t_d^{\rm min}\sim
20$Myr. As discussed in Ref.~\citep{Dominik:2013}, the redshift
evolution of x of the host galaxy affects the merger rate of
DNS binaries (see their top panel of Figures 3 and 4.). For higher metallicities, the peak of the merger rate density shifts towards lower redshifts. From this perspective, our M$_{\rm{Wilkins}}$ is representative of the case where  the DNS mergers dominantly  happen in high metallicity environments, shifting the peak towards lower redshifts.
	We also consider another model of rate density evolution, obtained by \cite{WP15}, 
	\[ R_{WP}(z) = 45 Mpc^{-3} Gyr^{-1}.
	\left\{
	\begin{array}{ll}
	e^{(z-0.9)/0.39} &  z \leq 0.9 \\
	e^{-(z-0.9)/0.26} &  z > 0.9
	\end{array} \right. \] \label{eq-short-z}
	This is a model (denoted as M$_{\text{WP}}$) derived based on the short GRBs observed by the Gamma-Ray satellites accounting for the effect of beaming. Though, some what indirect, we use this model to have enough diversity in the set of models we compare against.

	Along with these models, we also consider a case with constant comoving rate density $M_0$ characterized by $R(z)=R_0=1 Mpc^{-3} Myr^{-1}$. Left panel of  Fig.~\ref{fig:various} shows the normalized forms of all the merger rate density models discussed above.

	Given the merger rate density $R(z)$ (in units of $Mpc^{-3} Myr^{-1}$), the total number of sources (in units of $Myr^{-1}$) in a co-moving volume of radius $D$ is, 
	\begin{equation}
		N(D)\propto\int_0^{D}\frac{R(z(D'))}{1+z(D')}D'^2 dD'
	\label{total-no},
	\end{equation}
	where $z$ can be numerically inverted to obtain the corresponding co-moving distance $D$. The $(1+z)$ factor in the denominator accounts for the time dilation between the source-frame and the observer-frame. 
	
	Considering the proposed models above to be the underlying
source distribution within the co-moving volume and assuming isotropy,
we obtain the  optimal SNR distribution of DNS mergers for CE (right
panel of Fig.~\ref{fig:various}). To generate the source population for obtaining the optimal SNR distributions corresponding to different $R(z)$ (left panel of Fig~\ref{fig:various}), first the sources are assumed to be uniformly located and oriented on the {sphere parametrized by the comoving distance}. This is achieved by making sure that the azimuth angles $\phi,\psi$ are drawn from a uniform distribution $\left[0,2\pi\right]$ and the  polar angles  $\theta,\iota$ are chosen such that their cosines are uniformly distributed between $\left[-1,1\right]$. These choices ensure that at any radius, the source population is uniformly located and oriented on the surface of the sphere. Further, we need to distribute the sources {\it within} the detection volume specified by the maximum radius $D_{max} (\rho_{th}, M)$ (or $z_{max}(\rho_{th},M)$), which depends on the SNR threshold (=12). Hence we choose $N(D)$ to be uniformly distributed between $N(1)$ and $N(D_{max})$ and for each realization, we numerically solve Eq~(\ref{total-no}) to obtain corresponding $D$ value from $N(D)$.

	Using the procedure outlined above, we compute the optimal SNR distributions for the different models by imposing the SNR threshold of 12 which is shown in the right panel of Fig~\ref{fig:various}. Next we discuss how many detections are required for these models to be statistically distinguishable from each other.

\section{Statistical tests of distinguishability of various models}\label{Statistical_tests}


	In this section, we demonstrate the distinguishability of the SNR distributions corresponding to all the merger rate density evolution models discussed in the previous section. First of all, we discuss how to account for the error bars on the SNRs associated with GW detections and then we discuss the distinguishability of different models.

\subsection{Error bars on the SNRs}\label{sec-error}
	In reality, GW detections are made applying certain detection threshold on the matched filter SNRs which are calculated by matching the observed data and a template bank of precomputed GW waveforms. However, the SNR distributions in the right panel of Fig.~\ref{fig:various} are produced using optimal SNR for each source, where optimal SNR is a point estimate (SNR of the best fit template from matched filtering). Therefore it is important to fold in the error bars to the point estimates of SNRs to account for the usage of matched-filtering in the process of GW detections. 

	Under the assumption of zero mean Gaussian random noise in the detectors, the matched filter SNR ($\sigma$) follows the Rice distribution~\citep{Rice45} of the following form 
	\begin{equation}\label{Ricean}
		f(\sigma,\rho)= \sigma\, \exp\Bigg(-\frac{\sigma^2+\rho^2}{2}\Bigg)\,I_0(\rho \sigma),
	\end{equation}
	where $\rho$ is the optimal SNR and $I_0$ is the zeroth-order modified Bessel function of the first kind (see \citep{Moore:2019pke} for a detailed discussion).
	In order to account for the errors on SNRs in our distributions, we first calculate the optimal SNR (say $\rho_i$) for each source and then replace it with a number chosen at random from the distribution $f(\sigma,\rho_i)$ ( Eq.~\ref{Ricean}).

\subsection{Statistical tests}

	Now we quantify the distinguishability of the different SNR distributions by employing the Anderson-Darling (AD)~\citep{Anderson1952} test. The AD test is a well-known tool used to assess whether a sample data belongs to a reference distribution. The test returns a probability value ($p$ value) for the ``null" hypothesis that {\it sample belongs to the reference distribution}. If the null hypothesis is true, the $p$ value distribution obtained by performing the experiment multiple time is uniform between 0 and 1 {with a median p value of 0.5}. If the sample does not belong to the reference distribution, the $p$ value distribution will sharply peak around 0. A \textit{p} value distribution weighted more towards 0, implies a stronger evidence of rejecting the null  hypothesis or ability to distinguish the two distributions.

\begin{figure*}
	\centering
	\includegraphics[scale=0.6]{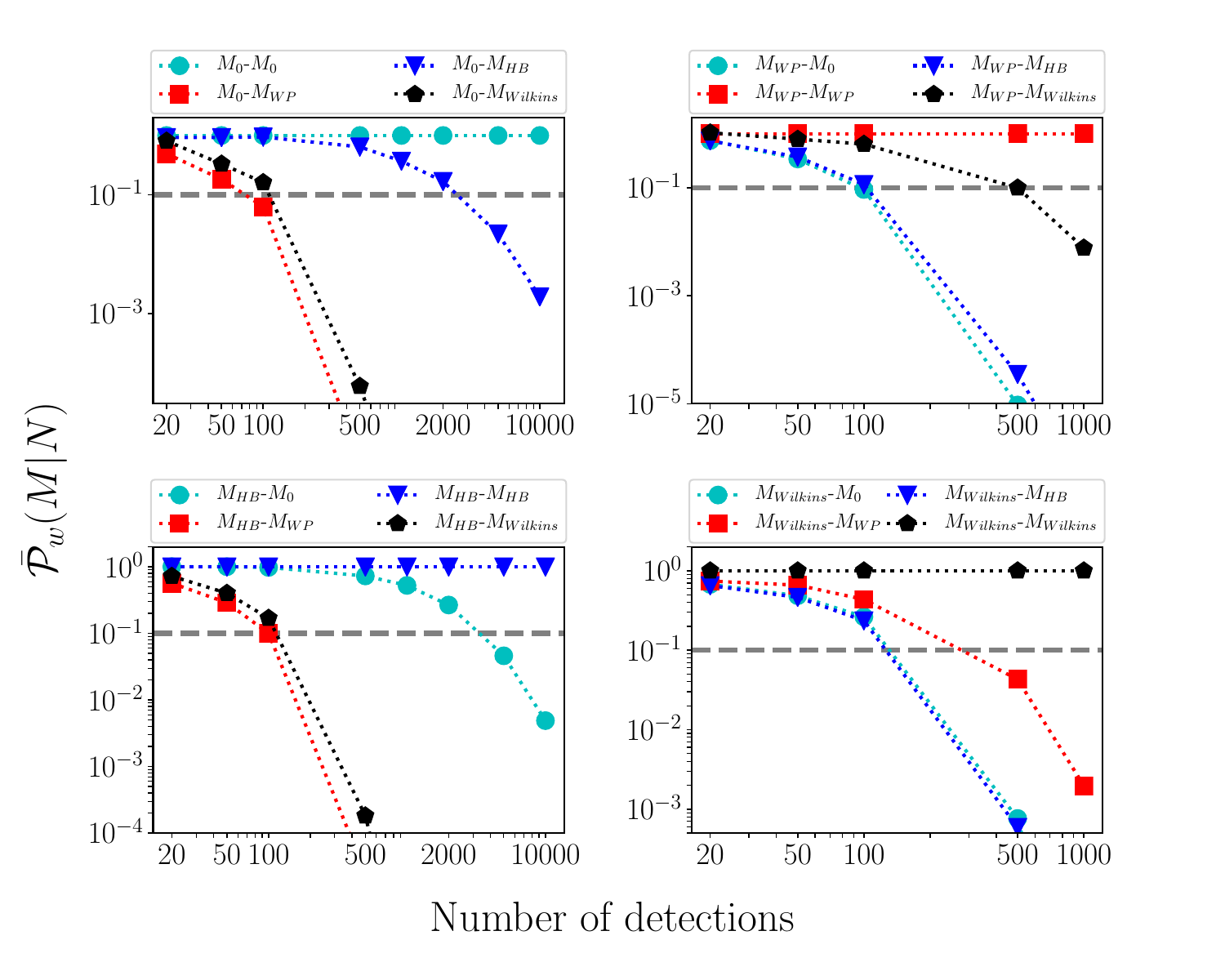}
	\caption{Weighted p-values (${\mathcal{\bar{P}}_w(M|N)}$) from Anderson-Darling test performed on the data obtained from the four models as function of number of detections. The first argument in each legend represents the data generated by following a particular model M (denoted as data$_M$ in the main text), where as the second argument is the theoretical model with which the data is compared to. We put a gray horizontal line in every panel corresponding to the threshold on ${\mathcal{\bar{P}}_w(M|N)}$ (see main text for  details).}\label{fig:p-value-stat}
\end{figure*}

	In order to quantify the distinguishability between two arbitrary merger rate density models M and N, we follow the procedure below. First we synthesize a fiducial data of SNR distribution of size \textit{n} (number of detections) assuming that the model M is the true distribution. As the data contains noise, we will account for the errors on the SNRs in the synthesized data along the lines mentioned earlier.  The \textit{data} is labelled as data$_M$, where the subscript refers to the underlying model. Then, we carry out the AD test between data$_{M}$ and the reference distribution $p_N(\rho)$ which is the predicted SNR distribution corresponding to the  model N. In the above step, since $p_N(\rho)$ is the theoretical prediction, it is always free of errors which is ensured by using sufficiently large number of samples to generate that. 
	
	The test returns a $p$ value which is denoted as ${\cal P}(M|N)$. Due to limited number (\textit{n}) of synthesized samples, the data$_M$ may not capture the essence of the model M and hence affects the \textit{p} value. 
	To  account for this, we repeat the \textit{p} value estimation 100 times, each time synthesizing the data$_M$ randomly and then compute the median of the resulting $p$ value distribution. The median of $p$ values is denoted as $\bar{\cal	P}(M|N)$. 


\subsection{Weighted $p$-values}
{	As mentioned earlier, we have used Rice distribution to
	model the errors in SNR. The presence of these errors in
	the data will affect the $p$ values which in turn can lead to false detection or false rejection.
	For example, the median of the $p$ value distribution resulting from
	AD test of data$_M$ with the model M, in principle,
	should be 0.5. But due to the errors, the test may return a lower median which may even lead
	to  the rejection of the null hypothesis when it is actually true. 
	In our case, we have multiple models \{N\} to be tested against the data$_M$
	and $p$ value for each model ($\bar{\cal{P}}(M|N)$) will decrease due to
	the errors thereby reducing the ability to distinguish between
	various models. } 

	{In order to quantify the distinguishability between the
	data and a model along the lines described earlier,  we introduce the
	notion of  \textit{weighted} $p$ values. For a
	given  data$_M$, we define a weighting function $\mathcal{W}$ as
	\beq
	\mathcal{W} = \frac{1}{\mathcal{\bar{P}}(M|M)}.
	\eeq  
	where ${\bar{P}}(M|M)$ is the median of $p$ values between
	data$_M$ and the underlying model $M$ (which, in the absence of noise,
	is 0.5). 
	We now define the weighted $p$ value between models data$_M$ and
	N as
	\begin{equation}
	\mathcal{\bar{P}}_w(M|N) = \mathcal{W}\,\times\,\mathcal{\bar{P}}(M|N).
	\end{equation}
}
The weighting factor $\mathcal{W}$ is chosen in such a way that
the weighted $p$ value for data$_M$ with the model $M$ itself always
returns unity ((\textit{i.e,} $\mathcal{\bar{P}}_w(M|M) = 1$)).
Weighted $p$ values have been extensively discussed in literature in
the context of testing multiple hypotheses (for example, see the
references \citep{Sture,benjamini2001,Guillermo}). {The  definition
	we use here may be thought of as an adaptation of this generic definition to our problem.}

Based on our previous discussion, it is clear that if two distributions are distinguishable, ${\mathcal{\bar{P}}_w(M|N)}$ will always be smaller than 1. Using a threshold on the median of the p-value distribution of 0.05  while performing the AD test (i.e. 95\% of the time the model is rejected), we set a rejection threshold on ${\mathcal{\bar{P}}_w(M|N)}$ to be 0.05/0.5=0.1.

\subsection{Effect of different SFR models on SNR distribution}

Here we discuss the distinguishability of different resulting SNR distributions from four different SFR models and  present our results in Fig.~\ref{fig:p-value-stat} where, in
the x-axis, we show the number of detections \textit{n} (for n= 20, 50,
100, 200, 500, 1000, 2000, 5000, 10000) and in the y-axis, we show the
distinguishability of each of the four rate models ($M_0,M_{\rm
HB},M_{\rm Wilkins},M_{\text{WP}}$) from each other by computing the
\textit{weighted} \textit{p} values ${\mathcal{\bar{P}}_w(M|N)}$ among
them. Each panel corresponds to a particular model for the data and the
different curves in each panel corresponds to
${\mathcal{\bar{P}}_w(M|N)}$ estimated for all the four models. For
example, in the top-left panel of Fig.~\ref{fig:p-value-stat} we
synthesize the data following constant co-moving rate density and compare against the theoretical distributions of the all four models. By
construction, the \textit{weighted} $p$ value
${\mathcal{\bar{P}}_w(M_0|M_0)}$, { when the data containing $M_0$
is compared with model M$_0$ itself}, represented by the cyan line, is
constant and is 1. As opposed to this scenario, all the other ratio
falls off as a function of the number of detections. Hence the data can
be distinguished from the other models. In the top-left panel, we also
see that a low number of detections ($\sim 500$) is sufficient to
distinguish between  data$_{M_0}$ and the model $M_{\text{WP}}$ or
$M_{\rm{Wilkins}}$ whereas we need at least thousands of detections to
differentiate between the data$_{M_0}$ and the model M$_\text{HB}$.

	{In the remaining three panels, we perform the same exercise for
the rest of the three models. In the top right panel the data
($\rm{data}_{M_{\rm{WP}}}$) is generated from the merger rate density
model, M$_{\rm WP}$. As expected, ${\mathcal{\bar{P}}_w(M_{\rm
WP}|M_{\rm WP})}$ is unity (red curve). The cyan, blue and the black
curves represent the comparison with models M$_0$, M$_\text{HB}$,
M$_{\rm Wilkins}$ respectively. We find that for a few hundreds of
detections, all the three models are distinguishable from M$_{\rm WP}$.
{This is not surprising given how different this distribution is from
others in the left panel of Fig.~\ref{fig:various}.}
		
	Similarly in the bottom left panel the data is generated
following the merger rate density model, M$_\text{HB}$ and in the bottom
right panel the data is generated following the merger rate density
model, M$_{\rm Wilkins}$. In case of data containing ${M_\text{HB}}$ (bottom
left panel) we find that larger number of detections ($\sim$ few thousands) of DNS
mergers are needed in order to distinguish this model especially from
M$_0$. As opposed to this scenario, in the bottom right panel, M$_{\rm
Wilkins}$ is distinguishable from other models given a few hundreds of detections.}
	
	Therefore it is evident that M$_{\rm WP}$ and M$_{Wilkins}$ can
be distinguished from all other models with high confidence with a few
hundreds of detections, whereas M$_{\text{HB}}$ is difficult to
distinguish from the $M_0$ using this method with low number of
detections. However, with large number of detections (say 10,000)
M$_{\text{HB}}$ is distinguishable from the other models. Given a sufficiently large number of detections, we expect  ${\mathcal{\bar{P}}_w(M|N)}$ to be either 0 or 1 given the two distributions are different or the same respectively. Hence we do not show any ${\mathcal{\bar{P}}_w(M|N)}<10^{-4}$ in Figure~\ref{fig:p-value-stat} and treat them as a scenario where the two distributions are completely distinguishable.

	As shown in \cite{ArunET09} and the most recent work by \cite{Baibhav:2019gxm}, the forecasted DNS detection rates by the third generation detectors ET-B and CE ranges from one thousand to tens of thousands per year. Given this rate, it is clear that the SNR data collected from less than an year of observation will be sufficient to test various merger rate density models. 


\begin{figure*}
	\includegraphics[scale=0.6]{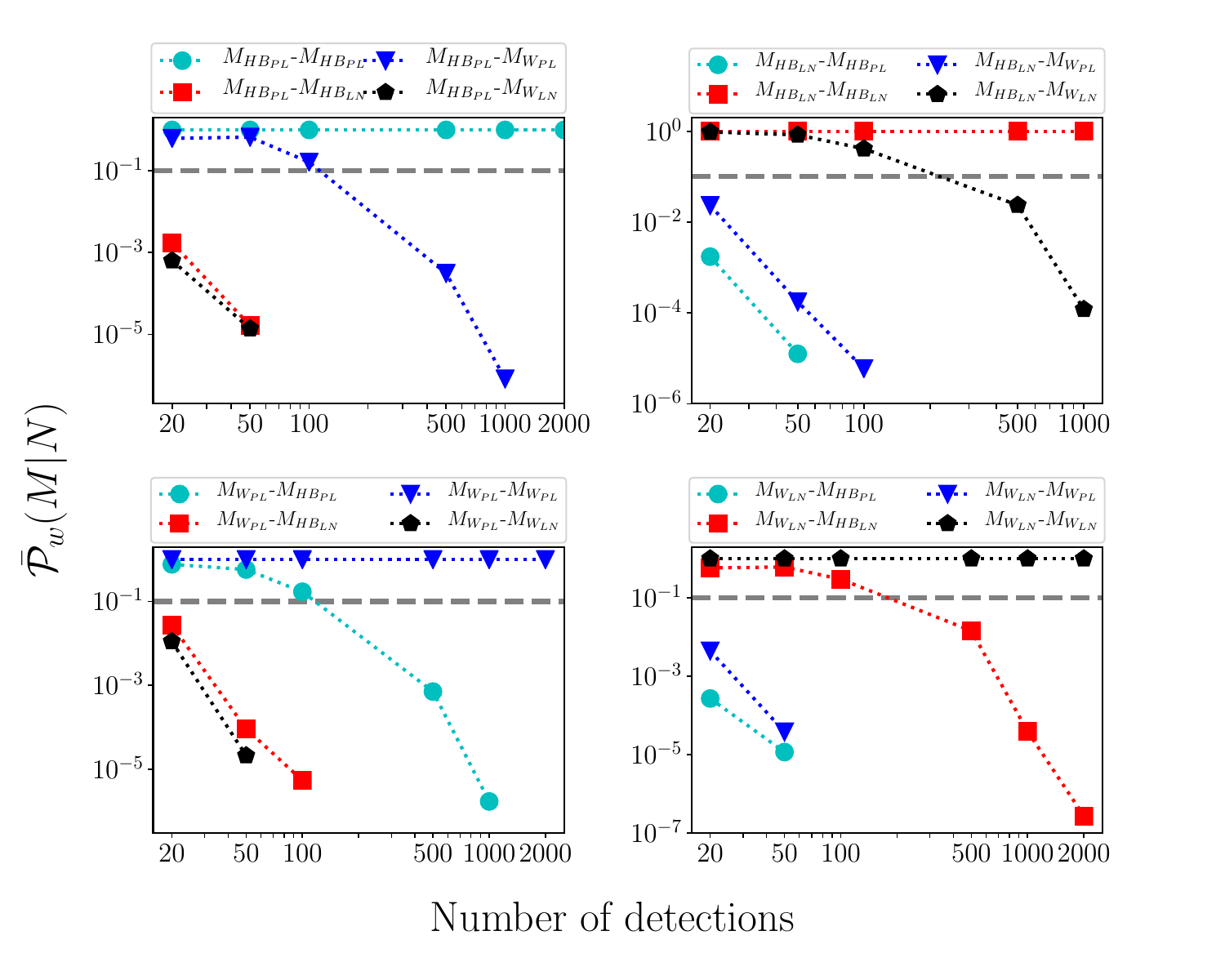}
		\caption{Same as Fig.~\ref{fig:p-value-stat} but
performed between models that are obtained from various combinations of
delay time distributions and star formation rate models. This is to
demonstrate how  different delay time models can be captured by the SNR
distributions and also to show how the uncertainty in delay time models
can contaminate the star formation models. { See main text for a
quantitative discussion.}}
		\label{fig:p-value-stat-delaytime}
\end{figure*}

\subsection{Effect of different delay time models on SNR distributions}

{In the previous section, we have shown the effects of the different merger rate distribution models on the observed SNR distributions. 
In this section, we investigate how the distribution of delay time affects the observed SNR distributions. 
As is evident from Eq.~\ref{eq;Rz}, the merger rate density is a convolution of 
star formation rate the delay time distribution. For the analysis in the
preceding section, we have used a power-law model for the delay time
distribution, $p(t_d) \propto {t_d}^{-1}$, with four different star formation rates. However, there are different models proposed in literature to represent the actual underlying delay time distribution. In this section, we investigate  the ability of SNR distributions to distinguish between  different delay-time models. } 


{We consider two models for delay time distributions. The first is a power-law (PL) distribution characterized by a power-law coefficient $\alpha_t$, 
\begin{equation}
p(t_d) \propto t_d^{-\alpha_t}
\label{eq:delay-powerlaw}
\end{equation}
and the second is a log-normal (LN)  distribution given as,
\begin{equation}
p(t_d) = \exp \left[  - \frac{(\ln  t_d - {\ln \tau})^2}{2 {\sigma_t}^2} \right]/\left( \sqrt{2\pi} \sigma_t \right),
\label{eq:delay-lognormal}
\end{equation}
where $\tau$ and $\sigma_t$ are model parameters.
These models have been considered in \cite{WP15} for estimating the short-GRB rate distribution. To be consistent with their estimated $1\sigma$ error bars on the model parameters, we have fixed  $\tau  = 3.9$ Gyr, $\sigma_t =  0.05$ and $\alpha_t = 0.81$ for our study. We have used these two models together with the two SFR models proposed in \cite{Hopkins_Beacom06} (HB) and \cite{Wilkins08} (W). Using Eq.~\ref{eq;Rz}, we obtain the resulting four different merger rate distributions labeled as  HB$_\text{PL}$, HB$_\text{LN}$, W$_\text{PL}$, W$_\text{LN}$. Following the same procedure as in the previous section, in Fig.~\ref{fig:p-value-stat-delaytime}, we show the distinguishability of the SNR distributions corresponding to these four models (see Fig.~\ref{fig:p-value-stat} and the corresponding texts for figure description).} 

{We find that for a given SFR, the two delay time models considered here are distinguishable with a few tens of events while for a given delay time model,  the two SFR considered here require about hundred events to distinguish from each other. This feature could be specific to the particular models we considered here, but conveys the broader message that 
the SNR distributions are sensitive to the variations in delay time models and indicate that under the assumption of a given SFR model, the SNR distributions can be used to perform model selection between different delay time models and one can constrain their model parameters. 
Further, though not extensively studied in this work, the results also point to the possible  degenerate combinations of  SFR and delay time models which can give identical merger rates and hence identical SNR distributions. Therefore, it might be difficult to constrain both SFR and delay time models simultaneously unless we have large number of detected events as expected with third generation GW detectors.}

\subsection{Effect of sub-threshold events}\label{section:sub-threshold-evnets}
In Fig.~\ref{fig:p-value-stat}, we have seen that the distinguishability increases as the number of detected sources increases. For a given detector sensitivity, the number of detections can be increased only by lowering the threshold which means by the inclusion of sub-threshold events. However, unlike in Fig.~\ref{fig:p-value-stat}, the increased number of sources by lowering the threshold are associated with larger uncertainties as given by Eq.~\ref{Ricean} and this might result in reducing the discriminatory power of the detected distribution. In this section, we investigate this in detail to see which of these two complementary effects is dominant and hence to assess whether sub-threshold events might be of use. 

In order to demonstrate this, we repeat one of the tests shown in Fig.~\ref{fig:p-value-stat} using different detection  thresholds. To generate representative detections, we take 1000 sources distributed as per  the model $M_\text{HB}$ and apply the detection thresholds 8, 16, 32 and 64 to get the respective detected sources where the number of detections are highest for threshold 8 and lowest for 64. The result are shown in Fig.~\ref{fig:p-value-threshold} where these four data sets are tested against two models $M_\text{Wilkins}$ and  $M_\text{HB}$ which is  the underlying true model itself.  The black diamonds show the weighted p-values (${\mathcal{\bar{P}}_w(M|N)}$) for the data against the model $M_\text{Wilkins}$. We find that as we decrease the detection threshold, the median of the weighted p-values consistently decreases. This  shows that the discriminatory power between two models increases with the inclusion of more number sources coming from lowering the threshold despite their higher uncertainties.

In the exercise above, the data points are made up of only sources of astrophysical origin which follow an underlying distribution even at low thresholds. However, in a real scenario, things are more complicated as there will be many noise transients mimicking as signals which can contaminate the detected distributions. Lowering the threshold to extremely low values will make the detections dominated by false alarms and hence the test will fail to achieve the goal. Hence in order to take the advantage of sub-threshold events, one will have to choose the threshold by fine tuning the false alarm rates so that noise transients do not contaminate the detections.  

\begin{figure}
	\centering
	\includegraphics[scale=.54]{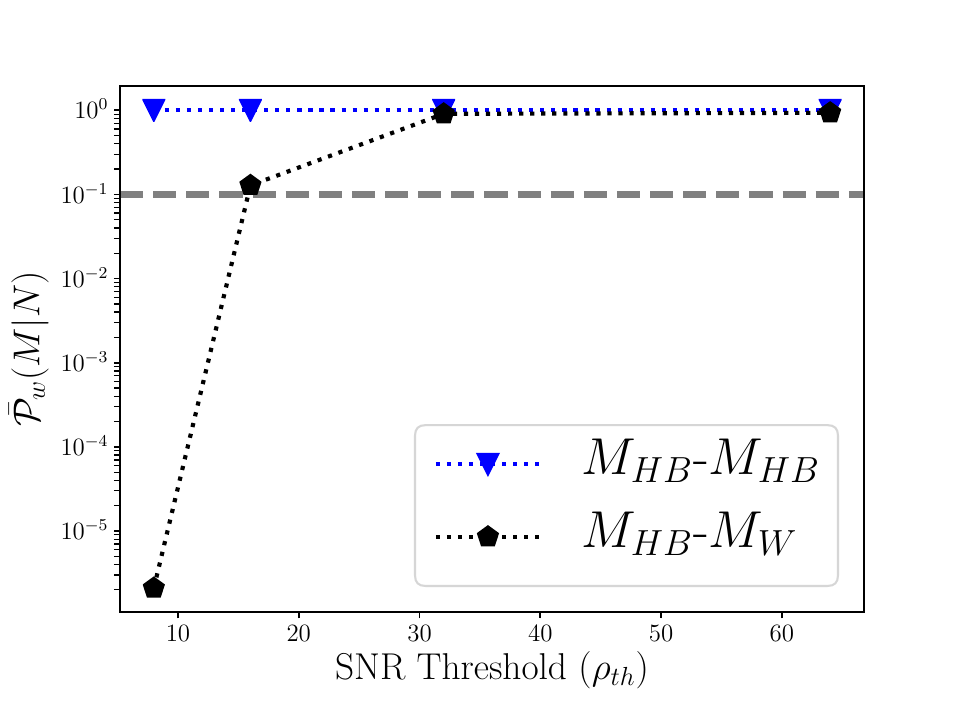}
	\caption{Weighted p-values (${\mathcal{\bar{P}}_w(M|N)}$)  shown as a function of the detection threshold. The data is generated as per the model $M_\text{HB}$ and tested against the model $M_\text{Wilkins}$ (shown as black diamonds). With lower detection thresholds, the Weighted p-value decreases which indicates higher discriminatory power by the inclusion of sub-threshold events. The blue triangles are the tests against same model as also shown in the previous figure.  We put a gray horizontal line corresponding to the threshold on ${\mathcal{\bar{P}}_w(M|N)}$ (see section~\ref{section:sub-threshold-evnets} for  details).}\label{fig:p-value-threshold}
\end{figure}


\section{Conclusions and Outlook}
{In this work, we have 
shown that the SNR distribution obtained from the gravitational wave detections of DNS mergers carries the imprints of merger rate evolution 
and hence can be used to  probe the redshift evolution of
binary neutron star mergers observed by third generation  gravitational wave observatories. 
Consequently, SNR distributions can also be used to probe the two ingredients of merger rates: the star formation rate model and the delay time distribution model for the time delay between the binary formation and merger. Unlike the traditional approach which uses luminosity distance estimates from the detailed follow-up studies of each detected DNS event, the proposed method only uses the signal to noise ratios of the DNS
mergers which are very likely to be available in real time by the time
the third generation detectors are operational. Hence a method based on
SNR distribution will be of interest. The importance of such a method is further emphasized in the context of
studies using the numerous sub-threshold events for which the parameter
estimation follow-up may not be performed (due to systematics~\cite{Huang:2018tqd}) while such events might still be of interest when there are sub-threshold triggers from EM counterpart searches in temporal and/or spatial coincidence. In fact it would be interesting to test if the inclusion of sub-threshold DNS triggers would alter our inference about  the merger rate evolution models obtained using only highly significant events (sub-threshold events are expected to have high redshifts - if they are astrophysical). 
 We also studied how many binary neutron star detections would be required to
rule in or rule out different known redshift evolution models. Using
Anderson-Darling test between the observed signal to noise ratio
distribution with various model predictions, we show that a few hundreds
of binary neutron star detections should be adequate to track the
redshift evolution of binary neutron stars. Further, our results indicate that the effects of star formation rates and delay time distributions can mimic each other and it might be challenging to constrain both from a given DNS merger 
distributions with a few hundreds of events. However, For a given SFR model assumption, a few tens to few hundred DNS detections will be enough to constrain the delay time models.}


\section{Acknowledgments}

K. G. A. and M. Saleem are partially supported by a grant
from Infosys Foundation. K. G. A. acknowledges the support
by the Indo-US Science and Technology Forum
through the Indo-US Center for the Exploration of
Extreme Gravity (Grant No. IUSSTF/JC-029/2016). We would like to thank Nathan Johnson-McDaniel, B. S. Sathyaprakash, P.
Ajith, Chris Van Den Broeck, Archisman Ghosh, Anuradha Gupta, Ghanashyam
Date, Alok Laddha and N V Krishnendu for very valuable discussions.
{We thank Maya
Fishbach for useful comments on an earlier version of the manuscript}. This document has LIGO preprint
number {\tt P1800002}.

\bibliographystyle{mnras}
\bibliography{Ref_list}

\begin{thebibliography}{}
\makeatletter
\relax
\def\mn@urlcharsother{\let\do\@makeother \do\$\do\&\do\#\do\^\do\_\do\%\do\~}
\def\mn@doi{\begingroup\mn@urlcharsother \@ifnextchar [ {\mn@doi@}
  {\mn@doi@[]}}
\def\mn@doi@[#1]#2{\def\@tempa{#1}\ifx\@tempa\@empty \href
  {http://dx.doi.org/#2} {doi:#2}\else \href {http://dx.doi.org/#2} {#1}\fi
  \endgroup}
\def\mn@eprint#1#2{\mn@eprint@#1:#2::\@nil}
\def\mn@eprint@arXiv#1{\href {http://arxiv.org/abs/#1} {{\tt arXiv:#1}}}
\def\mn@eprint@dblp#1{\href {http://dblp.uni-trier.de/rec/bibtex/#1.xml}
  {dblp:#1}}
\def\mn@eprint@#1:#2:#3:#4\@nil{\def\@tempa {#1}\def\@tempb {#2}\def\@tempc
  {#3}\ifx \@tempc \@empty \let \@tempc \@tempb \let \@tempb \@tempa \fi \ifx
  \@tempb \@empty \def\@tempb {arXiv}\fi \@ifundefined
  {mn@eprint@\@tempb}{\@tempb:\@tempc}{\expandafter \expandafter \csname
  mn@eprint@\@tempb\endcsname \expandafter{\@tempc}}}

\bibitem[\protect\citeauthoryear{Abbott et~al.}{Abbott
  et~al.}{2016a}]{BBH1strun}
Abbott B.~P.,  et~al., 2016a, \mn@doi [Phys. Rev.] {10.1103/PhysRevX.6.041015,
  10.1103/PhysRevX.8.039903}, X6, 041015

\bibitem[\protect\citeauthoryear{Abbott et~al.}{Abbott
  et~al.}{2016b}]{Discovery}
Abbott B.~P.,  et~al., 2016b, \mn@doi [Phys. Rev. Lett.]
  {10.1103/PhysRevLett.116.061102}, 116, 061102

\bibitem[\protect\citeauthoryear{Abbott et~al.}{Abbott et~al.}{2016c}]{TOG}
Abbott B.~P.,  et~al., 2016c, \mn@doi [Phys. Rev. Lett.]
  {10.1103/PhysRevLett.116.221101}, 116, 221101

\bibitem[\protect\citeauthoryear{Abbott et~al.}{Abbott et~al.}{2016d}]{PE}
Abbott B.~P.,  et~al., 2016d, \mn@doi [Phys. Rev. Lett.]
  {10.1103/PhysRevLett.116.241102}, 116, 241102

\bibitem[\protect\citeauthoryear{Abbott et~al.}{Abbott
  et~al.}{2016e}]{GW151226}
Abbott B.~P.,  et~al., 2016e, \mn@doi [Phys. Rev. Lett.]
  {10.1103/PhysRevLett.116.241103}, 116, 241103

\bibitem[\protect\citeauthoryear{Abbott et~al.}{Abbott et~al.}{2016f}]{Astro}
Abbott B.~P.,  et~al., 2016f, \mn@doi [Astrophys. J.]
  {10.3847/2041-8205/818/2/L22}, 818, L22

\bibitem[\protect\citeauthoryear{Abbott et~al.}{Abbott et~al.}{2016g}]{Rates}
Abbott B.~P.,  et~al., 2016g, \mn@doi [Astrophys. J.]
  {10.3847/2041-8205/833/1/L1}, 833, L1

\bibitem[\protect\citeauthoryear{Abbott et~al.}{Abbott et~al.}{2017a}]{CE_WB}
Abbott B.~P.,  et~al., 2017a, \mn@doi [Class. Quant. Grav.]
  {10.1088/1361-6382/aa51f4}, 34, 044001

\bibitem[\protect\citeauthoryear{Abbott et~al.}{Abbott
  et~al.}{2017b}]{GW170104}
Abbott B.~P.,  et~al., 2017b, \mn@doi [Phys. Rev. Lett.]
  {10.1103/PhysRevLett.118.221101}, 118, 221101

\bibitem[\protect\citeauthoryear{Abbott et~al.}{Abbott
  et~al.}{2017c}]{GW170814}
Abbott B.~P.,  et~al., 2017c, \mn@doi [Phys. Rev. Lett.]
  {10.1103/PhysRevLett.119.141101}, 119, 141101

\bibitem[\protect\citeauthoryear{Abbott et~al.}{Abbott
  et~al.}{2017d}]{GW170817}
Abbott B.,  et~al., 2017d, \mn@doi [Phys. Rev. Lett.]
  {10.1103/PhysRevLett.119.161101}, 119, 161101

\bibitem[\protect\citeauthoryear{Abbott et~al.}{Abbott
  et~al.}{2017e}]{LVCHubble}
Abbott B.~P.,  et~al., 2017e, \mn@doi [Nature] {10.1038/nature24471}, 551, 85

\bibitem[\protect\citeauthoryear{Abbott et~al.}{Abbott et~al.}{2017f}]{MMA}
Abbott B.~P.,  et~al., 2017f, \mn@doi [Astrophys. J.]
  {10.3847/2041-8213/aa91c9}, 848, L12

\bibitem[\protect\citeauthoryear{Abbott et~al.}{Abbott
  et~al.}{2017g}]{Multi-messenger2017}
Abbott B.~P.,  et~al., 2017g, \mn@doi [The Astrophysical Journal]
  {10.3847/2041-8213/aa91c9}, 848, L12

\bibitem[\protect\citeauthoryear{Abbott et~al.}{Abbott et~al.}{2017h}]{GW-GRB}
Abbott B.~P.,  et~al., 2017h, \mn@doi [Astrophys. J.]
  {10.3847/2041-8213/aa920c}, 848, L13

\bibitem[\protect\citeauthoryear{Abbott et~al.}{Abbott
  et~al.}{2017i}]{Kilonova_GW170817}
Abbott B.~P.,  et~al., 2017i, \mn@doi [The Astrophysical Journal]
  {10.3847/2041-8213/aa9478}, 850, L39

\bibitem[\protect\citeauthoryear{Abbott et~al.}{Abbott
  et~al.}{2017j}]{Progenitor-BNS}
Abbott B.~P.,  et~al., 2017j, \mn@doi [The Astrophysical Journal]
  {10.3847/2041-8213/aa93fc}, 850, L40

\bibitem[\protect\citeauthoryear{Abbott et~al.}{Abbott
  et~al.}{2017k}]{BNS-Postmerger}
Abbott B.~P.,  et~al., 2017k, \mn@doi [The Astrophysical Journal]
  {10.3847/2041-8213/aa9a35}, 851, L16

\bibitem[\protect\citeauthoryear{Abbott et~al.}{Abbott
  et~al.}{2017l}]{GW170608}
Abbott B.~P.,  et~al., 2017l, \mn@doi [Astrophys. J.]
  {10.3847/2041-8213/aa9f0c}, 851, L35

\bibitem[\protect\citeauthoryear{Abbott et~al.}{Abbott
  et~al.}{2018b}]{GW-Catalogue2}
Abbott B.~P.,  et~al., 2018b, preprint (\mn@eprint {arXiv} {1811.12940})

\bibitem[\protect\citeauthoryear{Abbott et~al.}{Abbott et~al.}{2018a}]{GWTC-1}
Abbott B.~P.,  et~al., 2018a, preprint (\mn@eprint {arXiv} {1811.12907})

\bibitem[\protect\citeauthoryear{Abbott et~al.}{Abbott
  et~al.}{2018c}]{BNS-radii19}
Abbott B.~P.,  et~al., 2018c, \mn@doi [Phys. Rev. Lett.]
  {10.1103/PhysRevLett.121.161101}, 121, 161101

\bibitem[\protect\citeauthoryear{Abbott et~al.}{Abbott
  et~al.}{2019}]{BNS-properties19}
Abbott B.~P.,  et~al., 2019, \mn@doi [Phys. Rev.] {10.1103/PhysRevX.9.011001},
  X9, 011001

\bibitem[\protect\citeauthoryear{Abernathy et~al.}{Abernathy
  et~al.}{2010}]{ArunET09}
Abernathy M.,  et~al., 2010, Einstein gravitational wave Telescope: Conceptual
  Design Study (Document number ET-0106A-10)

\bibitem[\protect\citeauthoryear{Ade et~al.}{Ade et~al.}{2014}]{Planck2013}
Ade P. A.~R.,  et~al., 2014, \mn@doi [Astron. Astrophys.]
  {10.1051/0004-6361/201321591}, 571, A16

\bibitem[\protect\citeauthoryear{Aghanim et~al.}{Aghanim
  et~al.}{2018}]{Planck2018}
Aghanim N.,  et~al., 2018, preprint (\mn@eprint {arXiv} {1807.06209})

\bibitem[\protect\citeauthoryear{Albert et~al.}{Albert
  et~al.}{2017}]{Neutrinos-BNS}
Albert A.,  et~al., 2017, \mn@doi [The Astrophysical Journal]
  {10.3847/2041-8213/aa9aed}, 850, L35

\bibitem[\protect\citeauthoryear{Anderson \& Darling}{Anderson \&
  Darling}{1952}]{Anderson1952}
Anderson T.~W.,  Darling D.~A.,  1952, \mn@doi [Ann. Math. Statist.]
  {10.1214/aoms/1177729437}, 23, 193

\bibitem[\protect\citeauthoryear{Ando}{Ando}{2004}]{Ando_2004}
Ando S.,  2004, \mn@doi [Journal of Cosmology and Astroparticle Physics]
  {10.1088/1475-7516/2004/06/007}, 2004, 007

\bibitem[\protect\citeauthoryear{Armano et~al.}{Armano et~al.}{2016}]{LPF}
Armano M.,  et~al., 2016, \mn@doi [Phys. Rev. Lett.]
  {10.1103/PhysRevLett.116.231101}, 116, 231101

\bibitem[\protect\citeauthoryear{Aso, Michimura, Somiya, Ando, Miyakawa,
  Sekiguchi, Tatsumi  \& Yamamoto}{Aso et~al.}{2013}]{KAGRA_ref}
Aso Y.,  Michimura Y.,  Somiya K.,  Ando M.,  Miyakawa O.,  Sekiguchi T.,
  Tatsumi D.,   Yamamoto H.,  2013, \mn@doi [Phys. Rev.]
  {10.1103/PhysRevD.88.043007}, D88, 043007

\bibitem[\protect\citeauthoryear{{Babak} et~al.,}{{Babak}
  et~al.}{2017}]{Babak2017}
{Babak} S.,  et~al., 2017, preprint, \href
  {http://adsabs.harvard.edu/abs/2017arXiv170309722B} {} (\mn@eprint {arXiv}
  {1703.09722})

\bibitem[\protect\citeauthoryear{Baibhav, Berti, Gerosa, Mapelli, Giacobbo,
  Bouffanais  \& Di~Carlo}{Baibhav et~al.}{2019}]{Baibhav:2019gxm}
Baibhav V.,  Berti E.,  Gerosa D.,  Mapelli M.,  Giacobbo N.,  Bouffanais Y.,
  Di~Carlo U.~N.,  2019, preprint (\mn@eprint {arXiv} {1906.04197})

\bibitem[\protect\citeauthoryear{Belczynski \& Kalogera}{Belczynski \&
  Kalogera}{2001}]{Belczynski00}
Belczynski K.,  Kalogera V.,  2001, \mn@doi [Astrophys. J.] {10.1086/319641},
  550, L183

\bibitem[\protect\citeauthoryear{Belczynski, Taam, Rantsiou  \& van~der
  Sluys}{Belczynski et~al.}{2008}]{Belczynski_2008}
Belczynski K.,  Taam R.~E.,  Rantsiou E.,   van~der Sluys M.,  2008, \mn@doi
  [The Astrophysical Journal] {10.1086/589609}, 682, 474

\bibitem[\protect\citeauthoryear{Benjamini \& Yekutieli}{Benjamini \&
  Yekutieli}{2001}]{benjamini2001}
Benjamini Y.,  Yekutieli D.,  2001, \mn@doi [Ann. Statist.]
  {10.1214/aos/1013699998}, 29, 1165

\bibitem[\protect\citeauthoryear{{Chen} \& {Holz}}{{Chen} \&
  {Holz}}{2014}]{ChenHolz2014}
{Chen} H.-Y.,  {Holz} D.~E.,  2014, preprint, \href
  {http://adsabs.harvard.edu/abs/2014arXiv1409.0522C} {} (\mn@eprint {arXiv}
  {1409.0522})

\bibitem[\protect\citeauthoryear{Cutler \& Flanagan}{Cutler \&
  Flanagan}{1994}]{CF94}
Cutler C.,  Flanagan E.,  1994, Phys. Rev. D, 49, 2658

\bibitem[\protect\citeauthoryear{D'Avanzo et~al.}{D'Avanzo
  et~al.}{2018}]{DAvanzo:2018zyz}
D'Avanzo P.,  et~al., 2018, \mn@doi [Astron. Astrophys.]
  {10.1051/0004-6361/201832664}, 613, L1

\bibitem[\protect\citeauthoryear{Dominik, Belczynski, Fryer, Holz, Berti,
  Bulik, Mandel  \& O'Shaughnessy}{Dominik et~al.}{2013}]{Dominik:2013}
Dominik M.,  Belczynski K.,  Fryer C.,  Holz D.~E.,  Berti E.,  Bulik T.,
  Mandel I.,   O'Shaughnessy R.,  2013, \mn@doi [Astrophys. J.]
  {10.1088/0004-637X/779/1/72}, 779, 72

\bibitem[\protect\citeauthoryear{{Durand}}{{Durand}}{2017}]{Guillermo}
{Durand} G.,  2017, arXiv e-prints, \href
  {https://ui.adsabs.harvard.edu/abs/2017arXiv171001094D} {p. arXiv:1710.01094}

\bibitem[\protect\citeauthoryear{Dwyer, Sigg, Ballmer, Barsotti, Mavalvala  \&
  Evans}{Dwyer et~al.}{2015}]{CEDwyer}
Dwyer S.,  Sigg D.,  Ballmer S.~W.,  Barsotti L.,  Mavalvala N.,   Evans M.,
  2015, \mn@doi [Phys. Rev. D] {10.1103/PhysRevD.91.082001}, 91, 082001

\bibitem[\protect\citeauthoryear{Sathyaprakash et~al.}{ET}{}]{ET}
{http://www.et-gw.eu/}

\bibitem[\protect\citeauthoryear{Fishbach, Holz  \& Farr}{Fishbach
  et~al.}{2018}]{FishbachHolz2018}
Fishbach M.,  Holz D.~E.,   Farr W.~M.,  2018, \mn@doi [Astrophys. J.]
  {10.3847/2041-8213/aad800}, 863, L41

\bibitem[\protect\citeauthoryear{Goldstein et~al.}{Goldstein
  et~al.}{2017}]{Goldstein:2017mmi}
Goldstein A.,  et~al., 2017, \mn@doi [Astrophys. J.]
  {10.3847/2041-8213/aa8f41}, 848, L14

\bibitem[\protect\citeauthoryear{Helstr\"om}{Helstr\"om}{1968}]{Helstrom68}
Helstr\"om C.,  1968, Statistical Theory of Signal Detection, 2nd edn.
 International Series of Monographs in Electronics and Instrumentation Vol. 9,
  Pergamon Press, Oxford, U.K., New York, U.S.A.

\bibitem[\protect\citeauthoryear{Hogg}{Hogg}{1999}]{Hogg1999}
Hogg D.~W.,  1999, preprint (\mn@eprint {arXiv} {astro-ph/9905116})

\bibitem[\protect\citeauthoryear{Holm}{Holm}{1979}]{Sture}
Holm S.,  1979, Scandinavian Journal of Statistics, 6, 65

\bibitem[\protect\citeauthoryear{{Hopkins} \& {Beacom}}{{Hopkins} \&
  {Beacom}}{2006}]{Hopkins_Beacom06}
{Hopkins} A.~M.,  {Beacom} J.~F.,  2006, \mn@doi [apj] {10.1086/506610}, \href
  {https://ui.adsabs.harvard.edu/abs/2006ApJ...651..142H} {651, 142}

\bibitem[\protect\citeauthoryear{Huang, Middleton, Ng, Vitale  \& Veitch}{Huang
  et~al.}{2018}]{Huang:2018tqd}
Huang Y.,  Middleton H.,  Ng K. K.~Y.,  Vitale S.,   Veitch J.,  2018, \mn@doi
  [Phys. Rev.] {10.1103/PhysRevD.98.123021}, D98, 123021

\bibitem[\protect\citeauthoryear{Iyer et~al.}{Iyer et~al.}{2011}]{Ligo-india}
Iyer B.,  et~al., 2011, LIGO-India Technical Report No. LIGO-M1100296

\bibitem[\protect\citeauthoryear{Lazzati, Perna, Morsony, López-Cámara,
  Cantiello, Ciolfi, Giacomazzo  \& Workman}{Lazzati
  et~al.}{2018}]{Lazzati:2017zsj}
Lazzati D.,  Perna R.,  Morsony B.~J.,  López-Cámara D.,  Cantiello M.,
  Ciolfi R.,  Giacomazzo B.,   Workman J.~C.,  2018, \mn@doi [Phys. Rev. Lett.]
  {10.1103/PhysRevLett.120.241103}, 120, 241103

\bibitem[\protect\citeauthoryear{{Lipunov}, {Postnov}, {Prokhorov}, {Panchenko}
   \& {Jorgensen}}{{Lipunov} et~al.}{1995}]{Lipunov95}
{Lipunov} V.~M.,  {Postnov} K.~A.,  {Prokhorov} M.~E.,  {Panchenko} I.~E.,
  {Jorgensen} H.~E.,  1995, \mn@doi [\apj] {10.1086/176512}, \href
  {https://ui.adsabs.harvard.edu/abs/1995ApJ...454..593L} {454, 593}

\bibitem[\protect\citeauthoryear{Lyman et~al.}{Lyman
  et~al.}{2018}]{Lyman:2018qjg}
Lyman J.~D.,  et~al., 2018, \mn@doi [Nat. Astron.] {10.1038/s41550-018-0511-3},
  2, 751

\bibitem[\protect\citeauthoryear{Maggiore}{Maggiore}{2007}]{Maggiore}
Maggiore M.,  2007, {Gravitational Waves. Vol. 1: Theory and Experiments}.
Oxford Master Series in Physics, Oxford University Press

\bibitem[\protect\citeauthoryear{Margutti et~al.}{Margutti
  et~al.}{2018}]{Margutti:2018xqd}
Margutti R.,  et~al., 2018, \mn@doi [Astrophys. J.] {10.3847/2041-8213/aab2ad},
  856, L18

\bibitem[\protect\citeauthoryear{Moore, Gerosa  \& Klein}{Moore
  et~al.}{2019}]{Moore:2019pke}
Moore C.~J.,  Gerosa D.,   Klein A.,  2019, preprint (\mn@eprint {arXiv}
  {1905.11998})

\bibitem[\protect\citeauthoryear{O'Shaughnessy, Kalogera  \&
  Belczynski}{O'Shaughnessy et~al.}{2008}]{OShaughnessy07}
O'Shaughnessy R.~W.,  Kalogera V.,   Belczynski K.,  2008, \mn@doi [Astrophys.
  J.] {10.1086/526334}, 675, 566

\bibitem[\protect\citeauthoryear{Regimbau \& Hughes}{Regimbau \&
  Hughes}{2009}]{Regimbau09}
Regimbau T.,  Hughes S.~A.,  2009, \mn@doi [Phys. Rev.]
  {10.1103/PhysRevD.79.062002}, D79, 062002

\bibitem[\protect\citeauthoryear{Resmi et~al.}{Resmi
  et~al.}{2018}]{Resmi:2018wuc}
Resmi L.,  et~al., 2018, \mn@doi [Astrophys. J.] {10.3847/1538-4357/aae1a6},
  867, 57

\bibitem[\protect\citeauthoryear{Rice}{Rice}{1945}]{Rice45}
Rice S.~O.,  1945, Bell System Technical Journal, 24, 46

\bibitem[\protect\citeauthoryear{Ruan, Nynka, Haggard, Kalogera  \& Evans}{Ruan
  et~al.}{2018}]{Ruan:2017bha}
Ruan J.~J.,  Nynka M.,  Haggard D.,  Kalogera V.,   Evans P.,  2018, \mn@doi
  [Astrophys. J.] {10.3847/2041-8213/aaa4f3}, 853, L4

\bibitem[\protect\citeauthoryear{Sathyaprakash \& Schutz}{Sathyaprakash \&
  Schutz}{2009}]{SathyaSchutzLivRev09}
Sathyaprakash B.,  Schutz B.,  2009, Living Rev.Rel., 12, 2

\bibitem[\protect\citeauthoryear{Sathyaprakash et~al.}{Sathyaprakash
  et~al.}{2012}]{ET:Sathya}
Sathyaprakash B.,  et~al., 2012, \mn@doi [Class. Quant. Grav.]
  {10.1088/0264-9381/29/12/124013, 10.1088/0264-9381/30/7/079501}, 29, 124013

\bibitem[\protect\citeauthoryear{{Schutz}}{{Schutz}}{2011}]{Schutz2011}
{Schutz} B.~F.,  2011, \mn@doi [Classical and Quantum Gravity]
  {10.1088/0264-9381/28/12/125023}, \href
  {http://adsabs.harvard.edu/abs/2011CQGra..28l5023S} {28, 125023}

\bibitem[\protect\citeauthoryear{Troja et~al.,}{Troja
  et~al.}{2018}]{Troja:2018ruz}
Troja E.,  et~al., 2018, \mn@doi [Mon. Not. Roy. Astron. Soc.]
  {10.1093/mnrasl/sly061}, 478, L18

\bibitem[\protect\citeauthoryear{{Tutukov} \& {Yungelson}}{{Tutukov} \&
  {Yungelson}}{1994}]{Tutukov94}
{Tutukov} A.~V.,  {Yungelson} L.~R.,  1994, \mn@doi [mnras]
  {10.1093/mnras/268.4.871}, \href
  {https://ui.adsabs.harvard.edu/abs/1994MNRAS.268..871T} {268, 871}

\bibitem[\protect\citeauthoryear{Abbott et~al.}{VIR}{}]{VIRGO}
\url{http://www.virgo.infn.it}

\bibitem[\protect\citeauthoryear{Valenti et~al.,}{Valenti
  et~al.}{2017}]{KN-discovery-170817}
Valenti S.,  et~al., 2017, The Astrophysical Journal Letters, 848, L24

\bibitem[\protect\citeauthoryear{Van Den~Broeck}{Van
  Den~Broeck}{2014}]{Broeck13}
Van Den~Broeck C.,  2014, \mn@doi [J. Phys. Conf. Ser.]
  {10.1088/1742-6596/484/1/012008}, 484, 012008

\bibitem[\protect\citeauthoryear{Vitale}{Vitale}{2016}]{Vitale2016}
Vitale S.,  2016, \mn@doi [Phys. Rev. D] {10.1103/PhysRevD.94.121501}, 94,
  121501

\bibitem[\protect\citeauthoryear{Vitale, Farr, Ng  \& Rodriguez}{Vitale
  et~al.}{2018}]{Vitale:2018yhm}
Vitale S.,  Farr W.~M.,  Ng K.,   Rodriguez C.~L.,  2018

\bibitem[\protect\citeauthoryear{Wainstein \& Zubakov}{Wainstein \&
  Zubakov}{1962}]{Wainstein}
Wainstein L.~A.,  Zubakov V.~D.,  1962, Extraction of Signals from Noise.
Prentice-Hall, Englewood Cliffs

\bibitem[\protect\citeauthoryear{Wanderman \& Piran}{Wanderman \&
  Piran}{2015}]{WP15}
Wanderman D.,  Piran T.,  2015, \mn@doi [Mon. Not. Roy. Astron. Soc.]
  {10.1093/mnras/stv123}, 448, 3026

\bibitem[\protect\citeauthoryear{{Wilkins}, {Trentham}  \& {Hopkins}}{{Wilkins}
  et~al.}{2008}]{Wilkins08}
{Wilkins} S.~M.,  {Trentham} N.,   {Hopkins} A.~M.,  2008, in {Kodama} T.,
  {Yamada} T.,   {Aoki} K.,  eds,  Astronomical Society of the Pacific
  Conference Series Vol. 399, Panoramic Views of Galaxy Formation and
  Evolution. p.~178 (\mn@eprint {arXiv} {0803.4024})

\bibitem[\protect\citeauthoryear{Wright}{Wright}{2006}]{Wright:2006}
Wright E.~L.,  2006, \mn@doi [Publ. Astron. Soc. Pac.] {10.1086/510102}, 118,
  1711

\bibitem[\protect\citeauthoryear{de Freitas~Pacheco, Regimbau, Vincent  \&
  Spallicci}{de~Freitas~Pacheco et~al.}{2006}]{deFreitasPacheco:2005ub}
de Freitas~Pacheco J.~A.,  Regimbau T.,  Vincent S.,   Spallicci A.,  2006,
  \mn@doi [Int. J. Mod. Phys.] {10.1142/S0218271806007699}, D15, 235

\makeatother
\end{thebibliography}

\end{document}